\documentclass[final,3p,times]{elsarticle}

\usepackage{amssymb}
\usepackage[]{amsmath}
\usepackage{graphics}
\usepackage{amssymb}
\usepackage[]{amsmath}
\usepackage{amsthm}
\usepackage{setspace}
\usepackage{epsfig}
\usepackage{subfigure}
\usepackage{booktabs}

\usepackage{graphicx}

\makeatletter

\newcommand{\Rmnum}[1]{\expandafter\@slowromancap\romannumeral #1@}
\makeatother

\biboptions{numbers,sort&compress}
\usepackage[dvipdfm,colorlinks,linkcolor=red,anchorcolor=blue,citecolor=green]{hyperref} %

\usepackage{amsmath}
\usepackage{times}
\usepackage{anysize}
\marginsize{2.5cm}{2.5cm}{1cm}{1cm}

\selectfont   

\journal {}

\begin{document}

\begin{frontmatter}

\title{General $N$-Dark Soliton Solutions of the Multi-Component Mel'nikov System}

\author{Zhong Han$^{{\rm a}}$}

\author{Yong Chen$^{{\rm a,b}}$ \corref{cor1}}
\ead{ychen@sei.ecnu.edu.cn}

\cortext[cor1]{Corresponding author.}

\address{$^{{\rm a}}$Shanghai Key Laboratory of Trustworthy Computing, East China Normal University, Shanghai, 200062, People's Republic of China}
\address{$^{{\rm b}}$MOE International Joint Lab of Trustworthy Software, East China Normal University, Shanghai, 200062, People's Republic of China}

\begin{abstract}
A general form of $N$-dark soliton solutions of the multi-component Mel'nikov system is presented. Taking the coupled Mel'nikov system comprised of two-component short waves and one-component long wave as an example, its general $N$-dark-dark soliton solutions in Gram determinant form are constructed through the KP hierarchy reduction method.  The dynamics of single dark-dark soliton and two dark-dark solitons are discussed in detail. It can be shown that the collisions of dark-dark solitons are elastic and energies of the solitons in different components completely transmit through. In addition, the dark-dark soliton bound states including both stationary and moving cases are also investigated. An interesting feature for the coupled Mel'nikov system is that the stationary dark-dark soliton bound states can exist for all possible combinations of nonlinearity coefficients including all-positive, all-negative and mixed types, while the moving case are possible when they take opposite signs or they are both negative. The dynamics and several interesting structures of the solutions are illustrated through some figures.
\end{abstract}

\end{frontmatter}

\section{Introduction}
During the past decades, many studies have been done on coupled systems describing the interaction of short wave packets with long waves in nonlinear dispersive media as they are frequently used in the fields of plasma physics, fluid dynamics and solid state physics \cite{yoko1,yoko2,yoko3}. What's more, the nonlinear interaction of multiple waves results in several interesting new physical processes \cite{pre1,pre2,pre3}. On the other hand, the studies of multi-component nonlinear systems have received much attention in recent years \cite{gener1,gener2,gener3}. Of particular interest is the multi-component generalization of the nonlinear Schr\"{o}dinger (NLS) equation, the so-called vector NLS equation \cite{gener2,gener3,gener4,yan,ling1,ling2}. Another interesting example is the multi-component long-wave-short-wave resonance interaction (LSRI) system \cite{lak1,lak2,lak3}, or the so called multi-component Yajima-Oikawa (YO) system \cite{liu,chen1,chen2,chen3}.

In the current paper, we concentrate on the Mel'nikov system \cite{mel1,mel2,mel3,mel4}
\begin{align}
&\textmd{i} \Phi_y=\Phi_{xx}+u \Phi, \label{i01}\\
&u_{xt}+u_{xxxx}+3(u^2)_{xx}-3 u_{yy}+\sigma (\Phi \Phi^*)_{xx}=0,\label{i02}
\end{align}
where $\sigma=\pm 1$, $\Phi\equiv \Phi(x,y,t)$ is the complex short wave amplitude and $u\equiv u(x,y,t)$ is the real long wave amplitude, the subscripts denote partial differentiation and the asterisk means complex conjugate hereafter. This system is used to describe the interaction of long waves with short wave packets propagating on the $x$-$y$ plane at an angle to each other. It may be considered either as a generalization of the Kadomtsev-Petviashvili (KP) equation with the addition of a complex scalar field or as a generalization of the NLS equation with a real scalar field \cite{india}. Its two-component generalization consisting of two short wave components and one long wave component is given by
\begin{align}
&\textmd{i}\Phi^{(1)}_y=\Phi^{(1)}_{xx}+u \Phi^{(1)}, \label{i03}\\
&\textmd{i}\Phi^{(2)}_y=\Phi^{(2)}_{xx}+u \Phi^{(2)}, \label{i04}\\
&u_{xt}+u_{xxxx}+3(u^2)_{xx}-3 u_{yy}+(\sigma_1\Phi^{(1)} {\Phi^{(1)}}^*+\sigma_2\Phi^{(2)}{\Phi^{(2)}}^*)_{xx}=0,\label{i05}
\end{align}
where $\sigma_1,\sigma_2=\pm 1$. Actually, the above coupled system can be generalized to a multi-component system consisting of $M$ short wave components and one long wave component, which is written in the following form
\begin{align}
&\textmd{i}\Phi^{(k)}_y=\Phi^{(k)}_{xx}+u \Phi^{(k)},\ \ \ \ \ k=1,2,\cdots,M,\label{i06}\\
&u_{xt}+u_{xxxx}+3(u^2)_{xx}-3 u_{yy}+\Big(\sum^M_{k=1} \sigma_k\Phi^{(k)}{\Phi^{(k)}}^\ast \Big)_{xx}=0,\label{i07}
\end{align}
where $\sigma_k=\pm 1$ for $k=1,2,\cdots, M$.

The Mel'nikov system (\ref{i01})-(\ref{i02}) admits boomeron type solutions which can be realized from an asymptotic analysis of the two soliton solutions \cite{mel3}, and its multi-soliton solutions are derived in Ref. \cite{mel4} via the theory of matrices. Soliton solutions of bright- and dark-types have been obtained from the Wronskian solutions of the KP hierarchy equaions \cite{hase}. Its Painlev\'{e} property and some exponentially localized dromion type solutions are studied in Ref. \cite{india}. More recently, its rogue wave solutions are also derived by virtue of the Hirota's bilinear method \cite{qin}. However, as far as we know, the general multi-dark soliton solutions of the multi-component Mel'nikov system have not been reported yet. Actually, general multi-dark soliton solutions of two-dimensional multi-component integrable systems are rather rare excepted that the multi-dark soliton solutions of the two-dimensional (2D) multi-component Yajima-Oikawa (YO) systems are reported in Ref. \cite{chen1}.

The KP hierarchy reduction method for deriving soliton solutions of integrable systems is firstly developed by the Kyoto school \cite{jimbo1,jimbo2}, and later used to get solutions of the NLS equation, the modified KdV equation and the Davey-Stewartson (DS) equation. Indeed, the pseudo-reduction of the two-dimensional Toda lattice hierarchy to constrained KP systems with dark soliton solutions is established in Ref. \cite{will1}, while the reduction to constrained KP systems with bright soliton solutions from the multi-component KP hierarchy is developed in Ref. \cite{will2}. Using this reduction method, the general $N$-dark-dark solitons of a two-coupled focusing-defocusing NLS equations are obtained in Ref. \cite{ohta} by Ohta et al. Also by virtue of this method, the general bright-dark $N$-soliton solutions of the vector NLS equations with all possible combinations of nonlinearities containing all focusing, all defocusing and mixed types are constructed in Ref. \cite{feng} by Feng. More recently, this method has been applied to derive the $N$-dark soliton \cite{chen1} and mixed $N$-soliton \cite{chen2} solutions of multi-component YO system. In addition, the KP hierarchy reduction method has also been developed to construct rational solutions (lump and rogue wave solutions) of soliton equations \cite{ohta2,ohta3,ohta4}, see also Refs. \cite{chen3,shi}

In the present paper, the general $N$-dark-dark soliton solutions of Eqs.(\ref{i03})-(\ref{i05}) are obtained through the KP hierarchy reduction method and their dynamics are also studied in detail. Following the KP theory, we derive the general $N$-dark-dark soliton solutions in terms of Gram determinants from the $\tau$-function solutions of the KP hierarchy. For single dark-dark solitons, we show that the degrees of "darkness" in the two components are different in general. For the collisions of two dark-dark solitons, it can be shown that there is no energy transfer in two components of each soliton, hence they are elastic collisions. As a matter of fact, the soliton bound state is a fascinating subject in the soliton theory. The dark-dark soliton bound states in the 2D coupled YO system and the coupled NLS equations have been reported in Refs. \cite{chen1} and \cite{ohta} respectively. For those two models, it has been shown that the bound states exist only when the coefficients of nonlinear terms take opposite signs. Different from Refs. \cite{chen1} and \cite{ohta}, it can be shown that for the coupled Mel'nikov system (\ref{i03})-(\ref{i05}), the stationary dark-dark soliton bound states can exist for all possible combinations of nonlinearity coefficients including all-positive, all-negative and mixed types, while the moving case are possible when they take opposite signs or they are both negative.

The rest of this paper is organized as follows. In section 2, the general $N$-dark-dark soliton solutions in the Gram determinant form of the two-component Mel'nikov system are obtained via the KP hierarchy reduction method. Section 3 devotes to analysis the dynamics of single and two dark-dark soltions. In section 4, the dark-dark soliton bound states including both the stationary and the moving cases are investigated in detail. In section 5, the general $N$-dark soliton solutions of the multi-component Mel'nikov system are presented without a detail derivation. We summarize the paper in section 6.

\section{Dark-Dark Soliton Solutions of the Coupled Mel'nikov system}
By virtue of the dependent variable transformation
\begin{align}\label{i08}
& \Phi^{(1)}=\rho_1 {\rm e}^{\textmd{i}\theta_1} \frac{g}{f},\ \ \ \ \Phi^{(2)}=\rho_2 {\rm e}^{\textmd{i}\theta_2} \frac{h}{f},\ \ \ \ u=2 (\log f)_{xx},
\end{align}
where $f\equiv f(x,y,t)$ is a real function, $g\equiv g(x,y,t)$ and $h\equiv h(x,y,t)$ are two complex functions, $\rho_1$ and $\rho_2$ are two positive constants. Meanwhile, $\theta_i=\alpha_i x+\alpha^2_i y+\beta_i(t)$ for $i=1,2$, where $\alpha_i$ are real constants and $\beta_i(t)$ are arbitrary real functions. Then, Eqs.(\ref{i03})-(\ref{i05}) can be converted into the following bilinear forms
\begin{align}
& (D^2_x+2 \textmd{i} \alpha_1 D_x-{\rm i} D_y)g \cdot f=0,\label{i09}\\
& (D^2_x+2 \textmd{i} \alpha_2 D_x-{\rm i} D_y)h \cdot f=0,\label{i10}\\
& (D^4_x+D_xD_t-3D^2_y)f\cdot f=\sigma_1 \rho^2_1 (f^2-gg^\ast)+\sigma_2 \rho^2_2 (f^2-hh^\ast),\label{i11}
\end{align}
where $D$ is Hirota's bilinear differential operator defined as
\begin{align}\label{i12}
& D^l_xD^m_yD^n_tf(x,y,t)\cdot g(x,y,t)=\Big(\frac{\partial}{\partial x}-\frac{\partial}{\partial x'}\Big)^l\Big(\frac{\partial}{\partial y}-\frac{\partial}{\partial y'}\Big)^m\Big(\frac{\partial}{\partial t}-\frac{\partial}{\partial t'}\Big)^nf(x,y,t)\cdot g(x',y',t')|_{x=x',y=y',t=t'}.
\end{align}

If a new independent variable $s$ is introduced, Eq.(\ref{i11}) can be decoupled into
\begin{align}
& (D^4_x+D_xD_s-3D^2_y)f\cdot f=0,\label{i13}\\
& (D_xD_t-D_xD_s)f\cdot f=\sigma_1 \rho^2_1 (f^2-gg^\ast)+\sigma_2 \rho^2_2 (f^2-hh^\ast).\label{i14}
\end{align}

Following the KP theory, the equations under study are considered as a reduction of the KP hierarchy, then their soliton solutions in the Gram type or Wronski type determinants can be derived directly from the $\tau$ function of the KP hierarchy under this reduction. To this end, the Gram determinant solutions of the KP hierarchy are presented here.

\textbf{Lemma 1} The following bilinear equations in the KP hierarchy \cite{chen1,hase, ohta}
\begin{align}
& (D^2_{x_1}+2a D_{x_1}-D_{x_2})\tau(k+1,l)\cdot \tau(k,l)=0,\label{i15}\\
& \Big(\frac{1}{2}D_{x_1}D_{x_{-1}}-1\Big)\tau(k,l)\cdot \tau(k,l) =-\tau(k+1,l)\tau(k-1,l),\label{i16}\\
& (D^2_{x_1}+2b D_{x_1}-D_{x_2})\tau(k,l+1)\cdot \tau(k,l)=0,\label{i17}\\
& \Big(\frac{1}{2}D_{x_1}D_{y_{-1}}-1\Big)\tau(k,l)\cdot \tau(k,l) =-\tau(k,l+1)\tau(k,l-1),\label{i18}\\
& (D^4_{x_1}-4D_{x_1}D_{x_3}+3D^2_{x_2})\tau(k,l)\cdot \tau(k,l)=0,\label{i19}
\end{align}
where $a$ and $b$ are are complex constants, $k$ and $l$ are integers, have the Gram determinant solutions
\begin{align}\label{i20}
& \tau(k,l)=|m_{ij}(k,l)|_{1\leq i,j\leq N},
\end{align}
where the entries of the determinant are given by
\begin{align*}
& m_{ij}(k,l)=c_{ij}+\int \varphi_i(k,l) \psi_j(k,l) \textmd{d}x_1,\\
& \varphi_i(k,l)=(p_i-a)^k(p_i-b)^l \exp(\theta_{i}),\\
& \psi_j(k,l)=\Big(-\frac{1}{q_j+a}\Big)^k\Big(-\frac{1}{q_j+b}\Big)^l \exp(\tilde{\theta_{j}}),
\end{align*}
with
\begin{align*}
& \theta_i=\frac{1}{p_i-a}x_{-1}+\frac{1}{p_i-b}y_{-1}+p_ix_1+p^2_ix_2+p^3_ix_3+\theta_{i0},\\
& \tilde{\theta_j}=\frac{1}{q_j+a}x_{-1}+\frac{1}{q_j+b}y_{-1}+q_jx_1-q^2_jx_2+q^3_jx_3+\tilde{\theta}_{j0},
\end{align*}
where $c_{ij},p_i,q_j,\theta_{i0}$ and $\tilde{\theta}_{j0},(i,j=1,2,\cdots,N)$ are arbitrary complex constants.

Firstly, it is easy to verify that the $\varphi_i$ and $\psi_j$ given in Lemma 1 satisfy the following differential and difference relations
\begin{align}
& \partial_{x_2}\varphi_i(k,l)=\partial^2_{x_1}\varphi_i(k,l),\\
& \partial_{x_3}\varphi_i(k,l)=\partial^3_{x_1}\varphi_i(k,l),\\
& \varphi_i(k+1,l)=(\partial_{x_1}-a)\varphi_i(k,l),\\
& \partial_{x_2}\psi_i(k,l)=-\partial^2_{x_1}\psi_i(k,l),\\
& \partial_{x_3}\psi_i(k,l)=\partial^3_{x_1}\psi_i(k,l),\\
& \psi_i(k-1,l)=-(\partial_{x_1}+a)\psi_i(k,l).
\end{align}
Next, take the determinant
\begin{align*}
& \tau(k,l)=\det_{1\leq i,j\leq N}(m_{ij}(k,l))=|m_{ij}(k,l)|_{1\leq i,j\leq N},
\end{align*}
with the matrix elements $m_{ij}(k,l)$ defined as
\begin{align*}
& m_{ij}(k,l)=c_{ij}+\int \varphi_i(k,l) \psi_j(k,l) \textmd{d}x_1.
\end{align*}
Then it is easy to verify that $m_{ij}(k,l)$ satisfy
\begin{align}
& \partial_{x_{-1}}m_{ij}(k,l)=-\varphi_i(k-1,l)\psi_j(k+1,l),\\
& \partial_{x_{1}}m_{ij}(k,l)=\varphi_i(k,l)\psi_j(k,l),\\
& \partial_{x_{2}}m_{ij}(k,l)=(\partial_{x_1}\varphi_i(k,l))\psi_j(k,l)-\varphi_i(k,l)\partial_{x_1}\psi_j(k,l),\\
& \partial_{x_{3}}m_{i,j}(k,l)=(\partial^2_{x_1}\varphi_i(k,l))\psi_j(k,l)-(\partial_{x_1}\varphi_i(k,l))\partial_{x_1}\psi_j(k,l)+\varphi_i(k,l)\partial^2_{x_1}\psi_j(k,l),\\
& m_{ij}(k+1,l)=m_{ij}(k,l)+\varphi_i(k,l)\psi_j(k+1,l).
\end{align}

Note that Eqs.(\ref{i15}) and (\ref{i16}) are independent of the variable $x_3$, the differential and difference relations with respect to the variables $x_{-1},x_1$ and $x_2$ are the same with the Lemma 2.1 in Ref. \cite{chen1} and the Lemma 1 in Ref. \cite{ohta}, hence they hold true automatically. Here, we only need to prove that the determinant (\ref{i20}) satisfies the bilinear KP equation (\ref{i19}).

\textbf{Proof Lemma 1:} By virtue of the differential of determinant and the expansion formula of bordered determinant \cite{hirota}, the derivatives of the $\tau$ function can be expressed by the following bordered determinants
\begin{align}
& \partial_{x_{1}}\tau(k,l)=-\left| \begin{array}{ccccc}
m_{ij}(k,l) & \varphi_i(k,l)  \\
\psi_j(k,l) & 0
\end{array} \right|,\\
& \partial_{x_{2}}\tau(k,l)=-\left| \begin{array}{ccccc}
m_{ij}(k,l) & \partial_{x_{1}}\varphi_i(k,l)  \\
\psi_j(k,l) & 0
\end{array} \right|+\left| \begin{array}{ccccc}
m_{ij}(k,l) & \varphi_i(k,l)  \\
\partial_{x_{1}}\psi_j(k,l) & 0
\end{array} \right|,\\
& \partial_{x_{3}}\tau(k,l)=-\left| \begin{array}{ccccc}
m_{ij}(k,l) & \partial^2_{x_{1}}\varphi_i(k,l)  \\
\psi_j(k,l) & 0
\end{array} \right|+\left| \begin{array}{ccccc}
m_{ij}(k,l) & \partial_{x_{1}}\varphi_i(k,l)  \\
\partial_{x_{1}}\psi_j(k,l) & 0
\end{array} \right|-\left| \begin{array}{ccccc}
m_{ij}(k,l) & \varphi_i(k,l)  \\
\partial^2_{x_{1}}\psi_j(k,l) & 0
\end{array} \right|,\\
& \partial^2_{x_{1}}\tau(k,l)=-\left| \begin{array}{ccccc}
m_{ij}(k,l) & \varphi_i(k,l)  \\
\partial_{x_{1}}\psi_j(k,l) & 0
\end{array} \right|-\left| \begin{array}{ccccc}
m_{ij}(k,l) & \partial_{x_{1}}\varphi_i(k,l)  \\
\psi_j(k,l) & 0
\end{array} \right|,\\
& \partial^3_{x_{1}}\tau(k,l)=-\left| \begin{array}{ccccc}
m_{ij}(k,l) & \varphi_i(k,l)  \\
\partial^2_{x_{1}}\psi_j(k,l) & 0
\end{array} \right|-2 \left| \begin{array}{ccccc}
m_{ij}(k,l) & \partial_{x_{1}}\varphi_i(k,l)  \\
\partial_{x_{1}}\psi_j(k,l) & 0
\end{array} \right|-\left| \begin{array}{ccccc}
m_{ij}(k,l) & \partial^2_{x_{1}}\varphi_i(k,l)  \\
\psi_j(k,l) & 0
\end{array} \right|,\\
& \partial_{x_1}\partial_{x_{3}}\tau(k,l)=\left| \begin{array}{ccccc}
m_{ij}(k,l) & \partial^3_{x_{1}}\varphi_i(k,l)  \\
\psi_j(k,l) & 0
\end{array} \right|-\left| \begin{array}{ccccc}
m_{ij}(k,l) & \varphi_i(k,l)  \\
\partial^3_{x_{1}}\psi_j(k,l) & 0
\end{array} \right|-\left| \begin{array}{ccccc}
m_{ij}(k,l) & \varphi_i(k,l) &  \partial_{x_{1}}\varphi_i(k,l)\\
\psi_j(k,l) & 0 & 0\\
\partial_{x_{1}}\psi_j(k,l) & 0 & 0
\end{array} \right|,\\
\nonumber & \partial^2_{x_2}\tau(k,l)=\left| \begin{array}{ccccc}
m_{ij}(k,l) & \partial_{x_{1}}\varphi_i(k,l)  \\
\partial^2_{x_{1}}\psi_j(k,l) & 0
\end{array} \right|-\left| \begin{array}{ccccc}
m_{ij}(k,l) & \varphi_i(k,l)  \\
\partial^3_{x_{1}}\psi_j(k,l) & 0
\end{array} \right|+\left| \begin{array}{ccccc}
m_{ij}(k,l) & \partial^2_{x_{1}}\varphi_i(k,l)  \\
\partial_{x_{1}}\psi_j(k,l) & 0
\end{array} \right|\\
& \ \ \ \ \ \ \ \ \ \ \ \ \ \ -\left| \begin{array}{ccccc}
m_{ij}(k,l) & \partial^3_{x_{1}}\varphi_i(k,l)  \\
\psi_j(k,l) & 0
\end{array} \right|+2\left| \begin{array}{ccccc}
m_{ij}(k,l) & \varphi_i(k,l) &  \partial_{x_{1}}\varphi_i(k,l)\\
\psi_j(k,l) & 0 & 0\\
\partial_{x_{1}}\psi_j(k,l) & 0 & 0
\end{array} \right|,\\
\nonumber & \partial^4_{x_1}\tau(k,l)=-\left| \begin{array}{ccccc}
m_{ij}(k,l) & \varphi_i(k,l)  \\
\partial^3_{x_{1}}\psi_j(k,l) & 0
\end{array} \right|-\left| \begin{array}{ccccc}
m_{ij}(k,l) & \partial^3_{x_{1}}\varphi_i(k,l)  \\
\psi_j(k,l) & 0
\end{array} \right|-3\left| \begin{array}{ccccc}
m_{ij}(k,l) & \partial_{x_{1}}\varphi_i(k,l)  \\
\partial^2_{x_{1}}\psi_j(k,l) & 0
\end{array} \right|\\
& \ \ \ \ \ \ \ \ \ \ \ \ \ \ -3\left| \begin{array}{ccccc}
m_{ij}(k,l) & \partial^2_{x_{1}}\varphi_i(k,l)  \\
\partial_{x_{1}}\psi_j(k,l) & 0
\end{array} \right|+2\left| \begin{array}{ccccc}
m_{ij}(k,l) & \varphi_i(k,l) &  \partial_{x_{1}}\varphi_i(k,l)\\
\psi_j(k,l) & 0 & 0\\
\partial_{x_{1}}\psi_j(k,l) & 0 & 0
\end{array} \right|,
\end{align}
where the bordered determinants are defined as
\begin{align}
\left| \begin{array}{ccccc}
m_{ij} & \varphi_i  \\
\psi_j & 0
\end{array} \right|=\left| \begin{array}{ccccc}
m_{11} & m_{12} & \cdots & m_{1N} & \varphi_1  \\
m_{21} & m_{22} & \cdots & m_{2N} & \varphi_2  \\
\vdots & \vdots & \vdots & \vdots & \vdots \\
m_{N1} & m_{N2} & \cdots & m_{NN} & \varphi_N  \\
\psi_1 & \psi_2 & \cdots & \psi_N & 0
\end{array} \right|.
\end{align}

By virtue of the above expresses and the Jacobi formula of determinants, it is easy to verify that the $\tau$ function satisfies the bilinear KP equation (\ref{i19})
\begin{align}
\nonumber & (D^4_{x_1}-4D_{x_1}D_{x_3}+3D^2_{x_2})\tau(k,l) \cdot \tau(k,l)\\
\nonumber & =2\Big[\Big(\partial^4_{x_1}\tau(k,l)-4\partial_{x_1}\partial_{x_{3}}\tau(k,l)+3\partial^2_{x_2}\tau(k,l)\Big)\tau(k,l)-4\Big(\partial^3_{x_1}\tau(k,l)-\partial_{x_{3}}\tau(k,l)\Big)\partial_{x_1}\tau(k,l)\\
\nonumber & \ \ \ \ +3\Big(\partial^2_{x_1}\tau(k,l)-\partial_{x_{2}}\tau(k,l)\Big)\Big(\partial^2_{x_1}\tau(k,l)+\partial_{x_{2}}\tau(k,l)\Big)\Big]\\
\nonumber & = 24 \Big(\left| m_{ij}(k,l) \right| \cdot \left| \begin{array}{ccccc}
m_{ij}(k,l) & \varphi_i(k,l) &  \partial_{x_{1}}\varphi_i(k,l)\\
\psi_j(k,l) & 0 & 0\\
\partial_{x_{1}}\psi_j(k,l) & 0 & 0
\end{array} \right|+\left| \begin{array}{ccccc}
m_{ij}(k,l) & \varphi_i(k,l)  \\
\partial_{x_{1}} \psi_j(k,l) & 0
\end{array} \right| \cdot \left| \begin{array}{ccccc}
m_{ij}(k,l) & \partial_{x_{1}}\varphi_i(k,l)  \\
\psi_j(k,l) & 0
\end{array} \right|\\
\nonumber &  \ \ \ \ \ \ \ \ \ \ \  -\left| \begin{array}{ccccc}
m_{ij}(k,l) & \varphi_i(k,l)  \\
\psi_j(k,l) & 0
\end{array} \right| \cdot \left| \begin{array}{ccccc}
m_{ij}(k,l) & \partial_{x_{1}}\varphi_i(k,l)  \\
\partial_{x_{1}}\psi_j(k,l) & 0
\end{array} \right|\Big)\\
& =0.
\end{align}
This completes the proof of Lemma 1. $\Box$

Now we perform a reduction of the above bilinear equations to derive the general formulae for $N$-dark-dark soliton solutions of Eqs.(\ref{i03})-(\ref{i05}). Assuming $x_{-1},y_{-1},x_1,x_3$ are real, $x_2,a(=\textmd{i} \alpha_1),b(=\textmd{i} \alpha_2)$ are pure imaginary and $q_i=p^*_i,\tilde{\theta}_{j0}=\theta^*_{j0},c_{ji}=c^*_{ij}=\delta_{ij}$, then we can get
\begin{align}\label{i21}
& \tilde{\theta}_{j}=\theta^*_{j},\ \ \ \ m_{ji}(k,l)=m^*_{ij}(-k,-l),\ \ \ \ \tau(k,l)=\tau^*(-k,-l).
\end{align}
Moreover, by defining
\begin{align}\label{i22}
& f= \tau(0,0),\ \ \ \ g= \tau(1,0),\ \ \ \ h= \tau(0,1),\ \ \ \ g^*= \tau(-1,0),\ \ \ \ h^*= \tau(0,-1),
\end{align}
the bilinear equations (\ref{i15})-(\ref{i19}) can be recast into
\begin{align}
& (D^2_{x_1}+2\textmd{i}\alpha_1 D_{x_1}-D_{x_2})g\cdot f=0,\label{i23}\\
& \Big(\frac{1}{2}D_{x_1}D_{x_{-1}}-1\Big)f\cdot f =-gg^*,\label{i24}\\
& (D^2_{x_1}+2\textmd{i}\alpha_2 D_{x_1}-D_{x_2})h\cdot f=0,\label{i25}\\
& \Big(\frac{1}{2}D_{x_1}D_{y_{-1}}-1\Big)f\cdot f =-hh^*,\label{i26}\\
& (D^4_{x_1}-4D_{x_1}D_{x_3}+3D^2_{x_2})f\cdot f=0.\label{i27}
\end{align}

By virtue of the following independent variable transformation
\begin{align}\label{i28}
& x_1=x,\ \ \ \ x_2=-\textmd{i}y,\ \ \ \ x_3=-4(t+s), \ \ \ \ x_{-1}=\frac{1}{2}\sigma_1 \rho^2_1t, \ \ \ \ y_{-1}=\frac{1}{2}\sigma_2 \rho^2_2t,
\end{align}
the bilinear equations (\ref{i23})-(\ref{i27}) are recast into the bilinear form (\ref{i09})-(\ref{i10}) and (\ref{i13})-(\ref{i14}). Thus, applying the independent variable transformation (\ref{i28}) to the $f,g,h$ in (\ref{i22}) and neglecting the $s$ dependence, the following theorem for the general $N$-dark-dark soliton solutions of Eqs.(\ref{i03})-(\ref{i05}) is immediately obtained.

\textbf{Theorem 1} The $N$-dark-dark soliton solutions of the coupled Mel'nikov system (\ref{i03})-(\ref{i05}) are
\begin{align}
& \Phi^{(1)}=\rho_1 {\rm e}^{{\rm i}[\alpha_1 x+\alpha^2_1 y+\beta_1(t)]} \frac{g}{f},\label{i29}\\
&\Phi^{(2)}=\rho_2 {\rm e}^{{\rm i}[\alpha_2 x+\alpha^2_2 y+\beta_2(t)]} \frac{h}{f},\label{i30}\\
& u=2 (\log f)_{xx},\label{i31}
\end{align}
where $f,g$ and $h$ are Gram determinants given by
\begin{align*}
&f=\Bigg|  \delta_{ij} + \frac{1}{p_i+p^*_j} \textmd{e}^{\xi_i+\xi^*_j}  \Bigg|_{N\times N}, \\
&g=\Bigg| \delta_{ij} + \left(-\frac{p_i-\textmd{i}\alpha_1}{p^*_j+\textmd{i}\alpha_1}\right)  \frac{1}{p_i+p^*_j} \textmd{e}^{\xi_i+\xi^*_j} \Bigg|_{N\times N},\ \ \\
& h=\Bigg| \delta_{ij} + \left(-\frac{p_i-\textmd{i}\alpha_2}{p^*_j+\textmd{i}\alpha_2}\right)  \frac{1}{p_i+p^*_j} \textmd{e}^{\xi_i+\xi^*_j}  \Bigg|_{N\times N},\ \
\end{align*}
with
\begin{align*}
& \xi_j=p_jx -\textmd{i}p^2_jy
 + \frac{1}{2}\Big(\frac{\sigma_1\rho^2_1}{p_j-\textmd{i}\alpha_1}+\frac{\sigma_2\rho^2_2}{p_j-\textmd{i}\alpha_2}-8p^3_j \Big)t  + \xi_{j0},
\end{align*}
where $p_j$ and $\xi_{j0}$ are complex constants, $\delta_{ij}$ is the Kronecker symbol ($\delta_{ij}$ is 1 when $i=j$ and 0 otherwise).

\section{Dynamics of Dark-Dark Solitons}

\subsection{Single dark-dark solitons}
To get single dark-dark soliton solution of Eqs.(\ref{i01})-(\ref{i03}), we take $N=1$ in the formula (\ref{i29})-(\ref{i31}). The Gram determinants read
\begin{align}
&f=1 + \frac{1}{p_1+p^*_1} \textmd{e}^{\xi_1+\xi^*_1}, \label{i32}\\
&g=1-\frac{1}{p_1+p^*_1} \frac{p_1-\textmd{i}\alpha_1}{p^*_1+\textmd{i}\alpha_1}   \textmd{e}^{\xi_1+\xi^*_1}, \label{i33}\\
& h=1-\frac{1}{p_1+p^*_1} \frac{p_1-\textmd{i}\alpha_2}{p^*_1+\textmd{i}\alpha_2}   \textmd{e}^{\xi_1+\xi^*_1},\label{i34}
\end{align}
and the single dark-dark soliton solution can be written as
\begin{align}
& \Phi^{(1)}=\frac{\rho_1}{2} {\rm e}^{{\rm i}[\alpha_1 x+\alpha^2_1 y+\beta_1(t)]}\times \Big[1+K^{(1)}_1+(K^{(1)}_1-1) \tanh \Big(\frac{\xi_1+\xi^*_1+\Theta_1}{2}\Big)\Big],\label{i35}\\
&\Phi^{(2)}=\frac{\rho_2}{2} {\rm e}^{{\rm i}[\alpha_2 x+\alpha^2_2 y+\beta_2(t)]} \times \Big[1+K^{(2)}_1+(K^{(2)}_1-1) \tanh \Big(\frac{\xi_1+\xi^*_1+\Theta_1}{2}\Big)\Big],\label{i36}\\
& u=\frac{1}{2}(p_1+p^*_1)^2 \textmd{sech}^2\Big(\frac{\xi_1+\xi^*_1+\Theta_1}{2}\Big),\label{i37}
\end{align}
with
\begin{align*}
& \textmd{e}^{\Theta_1}=\frac{1}{p_1+p^*_1}=\frac{1}{2a_1},\\
& K^{(1)}_1=-\frac{p_1-\textmd{i}\alpha_1}{p^*_1+\textmd{i}\alpha_1}=-\frac{a_1+\textmd{i}(b_1-\alpha_1)}{a_1-\textmd{i}(b_1-\alpha_1)},\\
&  K^{(2)}_1=-\frac{p_1-\textmd{i}\alpha_2}{p^*_1+\textmd{i}\alpha_2}=-\frac{a_1+\textmd{i}(b_1-\alpha_2)}{a_1-\textmd{i}(b_1-\alpha_2)},\\
& \xi_1+\xi^*_1=2a_1x +4a_1b_1y
 + \Big[\frac{\sigma_1a_1\rho^2_1}{a^2_1+(b_1-\alpha_1)^2}+\frac{\sigma_2a_1\rho^2_2}{a^2_1+(b_1-\alpha_2)^2}-8a_1(a^2_1-3b^2_1) \Big]t  +2 \xi_{10R},
\end{align*}
where $p_1=a_1+\textmd{i}b_1$, and $a_1,b_1,\xi_{10R}$ are real constants, $\xi_{10R}$ is the real part of $\xi_{10}$.

From (\ref{i35})-(\ref{i37}), it is easy to know that the intensity functions $|\Phi^{(1)}|,|\Phi^{(2)}|$ of the short wave components and the long wave component $u$ moving at velocity $-\frac{1}{2}\Big(\frac{\sigma_1\rho^2_1}{a^2_1+(b_1-\alpha_1)^2}+\frac{\sigma_2\rho^2_2}{a^2_1+(b_1-\alpha_2)^2}\Big)+4(a^2_1-3b^2_1)$ along the $x$-direction. In addition, $|\Phi^{(1)}|\rightarrow \rho_1,|\Phi^{(2)}|\rightarrow \rho_2,u\rightarrow 0$ when $x,y\rightarrow \pm \infty$.

Taking $K^{(1)}_1=\exp(2\textmd{i}\phi^{(1)}_1)$ and $K^{(2)}_1=\exp(2\textmd{i}\phi^{(2)}_1)$, i.e., $2\phi^{(1)}_1$ and $2\phi^{(2)}_1$ are the phases of the constants $K^{(1)}_1$ and $K^{(2)}_1$ respectively. As $x$ and $y$ vary from $-\infty$ to $+\infty$, the phases of the short wave components $\Phi^{(1)}$ and $\Phi^{(2)}$ acquire shifts in the amount of $2\phi^{(1)}_1$ and $2\phi^{(2)}_1$ while the long wave component $u$ phase shifts is zero. Without loss of generality, we can restrict $- \pi< 2\phi^{(1)}_1,2\phi^{(2)}_1\leq \pi$, i.e., $-\frac{\pi}{2}< \phi^{(1)}_1,\phi^{(2)}_1\leq \frac{\pi}{2}$. Then at the center of the solitons where $\xi_1+\xi^*_1+\Theta_1=0$, intensities of the components are
$|\Phi^{(1)}|_{\textmd{center}}= \rho_1 \cos \phi^{(1)}_1,|\Phi^{(2)}|_{\textmd{center}}= \rho_2 \cos \phi^{(2)}_1,u=2a^2_1$. For the short wave components $\Phi^{(1)}$ and $\Phi^{(2)}$, the fact that the center intensities are lower than the background intensities $\rho_1$ and $\rho_2$, thus these solitons are dark-dark solitons. Further more, the intensity dips at the centers of the $\Phi^{(1)}$ and $\Phi^{(2)}$ components are controlled by the phase shifts $2\phi^{(1)}_1$ and $2\phi^{(2)}_1$ respectively, hence these phase shifts dictate how "dark" the center is.

According to the values of $\alpha_1$ and $\alpha_2$, there exist the following two different cases:

(1) $\alpha_1=\alpha_2$, then $K^{(1)}_1=K^{(2)}_1$, therefor $\phi^{(1)}_1=\phi^{(2)}_1$. For this case, the short wave components $\Phi^{(1)}$ and $\Phi^{(2)}$ are proportional to each other, and they have the same degrees of darkness at the center. In this situation, the dark-dark solitons of the coupled Mel'nikov system is equivalent to a scalar dark soliton of the single-component Mel'nikov system, thus it is viewed as a degenerate case similar to the coupled YO system \cite{chen1} and the coupled NLS equations \cite{ohta}. These degenerate solitons are illustrated in Fig.1, from which it can be seen that both the $\Phi^{(1)}$ and $\Phi^{(2)}$ components are zero intensity at the soliton center.


(2) $\alpha_1\neq \alpha_2$, then $K^{(1)}_1\neq K^{(2)}_1$, thus $\phi^{(1)}_1\neq \phi^{(2)}_1$. This suggests that the $\Phi^{(1)}$ and $\Phi^{(2)}$ components in these solitons are not proportional to each other, thus can not be reducible to scalar single dark soliton. In this non-degenerate case, the $\Phi^{(1)}$ and $\Phi^{(2)}$ components have different degrees of darkness at the center. As is shown in Fig.2, the intensity of the component $\Phi^{(1)}$ is zero intensity, while the intensity of the component $\Phi^{(2)}$ is nonzero intensity at its center.
%

\subsection{Two dark-dark solitons}
Two dark-dark solitons of Eqs.(\ref{i01})-(\ref{i03}) correspond to $N=2$ in the formula (\ref{i29})-(\ref{i31}). In this case, we have
\begin{align}
& \Phi^{(1)}=\rho_1 {\rm e}^{{\rm i}[\alpha_1 x+\alpha^2_1 y+\beta_1(t)]} \frac{g_2}{f_2},\label{i38}\\
&\Phi^{(2)}=\rho_2 {\rm e}^{{\rm i}[\alpha_2 x+\alpha^2_2 y+\beta_2(t)]} \frac{h_2}{f_2},\label{i39}\\
& u=2 (\log f_2)_{xx},\label{i40}
\end{align}
with
\begin{align}
& f_2=1+\textmd{e}^{\xi_1+\xi^*_1+\Theta_1}+\textmd{e}^{\xi_2+\xi^*_2+\Theta_2}+\Omega_{12}\textmd{e}^{\xi_1+\xi^*_1+\xi_2+\xi^*_2+\Theta_1+\Theta_2},\label{i41}\\
& g_2=1+K^{(1)}_1\textmd{e}^{\xi_1+\xi^*_1+\Theta_1}+K^{(1)}_2\textmd{e}^{\xi_2+\xi^*_2+\Theta_2}+\Omega_{12}K^{(1)}_1K^{(1)}_2\textmd{e}^{\xi_1+\xi^*_1+\xi_2+\xi^*_2+\Theta_1+\Theta_2},\label{i42}\\
& h_2=1+K^{(2)}_1\textmd{e}^{\xi_1+\xi^*_1+\Theta_1}+K^{(2)}_2\textmd{e}^{\xi_2+\xi^*_2+\Theta_2}+\Omega_{12}K^{(2)}_1K^{(2)}_2\textmd{e}^{\xi_1+\xi^*_1+\xi_2+\xi^*_2+\Theta_1+\Theta_2},\label{i43}
\end{align}
and
\begin{align*}
& \textmd{e}^{\Theta_j}=\frac{1}{p_j+p^*_j}=\frac{1}{2a_j},\\
& K^{(1)}_j=-\frac{p_j-\textmd{i}\alpha_1}{p^*_j+\textmd{i}\alpha_1}=-\frac{a_j+\textmd{i}(b_j-\alpha_1)}{a_j-\textmd{i}(b_j-\alpha_1)},\\
&  K^{(2)}_j=-\frac{p_j-\textmd{i}\alpha_2}{p^*_j+\textmd{i}\alpha_2}=-\frac{a_j+\textmd{i}(b_j-\alpha_2)}{a_j-\textmd{i}(b_j-\alpha_2)},\\
& \Omega_{12}=\Big|\frac{p_1-p_2}{p_1+p^*_2}\Big|^2=\frac{(a_1-a_2)^2+(b_1-b_2)^2}{(a_1+a_2)^2+(b_1-b_2)^2},\\
& \xi_j+\xi^*_j=2a_jx +4a_jb_jy
 + \Big[\frac{\sigma_1\rho^2_1a_j}{a^2_j+(b_j-\alpha_1)^2}+\frac{\sigma_2\rho^2_2a_j}{a^2_j+(b_j-\alpha_2)^2}-8a_j(a^2_j-3b^2_j) \Big]t  +2 \xi_{j0R}\\
& \ \ \ \ \ \ \ \ \ \ \   = k_{x,j}x+k_{y,j}y+\omega_jt+2 \xi_{j0R}
\end{align*}
where $p_j=a_j+\textmd{i}b_j$, $a_j,b_j,\xi_{j0R},(j=1,2)$ are real constants.

In the case of $a_2=-a_1$ and $b_2=b_1$, i.e., $p_2=-p^*_1$, the denominator of $\Omega_{12}$ is zero. The soliton interaction possess Y-shape on this particular wave number. This Y-shape type soliton solution is found in the KP equation and is also known as the resonant soliton solution. Analogous to the two-soliton solution of the KP equation, the above two soliton interaction solutions are classified into two different types \cite{kp1,kp2,kp3,kp4}:

(1) If $a_1a_2<0$, then $\Omega_{12}>1$. This case is called the O-type soliton interaction. In this situation, the two asymptotic soliton amplitudes $2a^2_1$
and $2a^2_2$ (in the long wave component $u$) are equivalent when $a_1=-a_2$. The interaction peak (the maximum of $u$) is always greater than the sum of the
asymptotic soliton amplitudes.

(2) If $a_1a_2>0$, then $0<\Omega_{12}<1$. This case is known as the P-type soliton interaction. In this situation, the two asymptotic soliton amplitudes $2a^2_1$
and $2a^2_2$ (also in $u$) cannot be equivalent. The interaction peak (the maximum of $u$) is always less than the sum of the asymptotic soliton amplitudes.

It is obvious that the types of soliton interaction do not depend on the parameters $b_1$ and $b_2$ (the imaginary parts of $p_1$ and $p_2$). The resonant Y-shape soliton solution can be obtained via taking the limit $b_2 \rightarrow b_1$ in the equal-amplitude O-type two-soliton solution ($a_1=-a_2$). In addition, the interaction coefficient $\Omega_{12}$ in the two-soliton solution of the coupled Mel'nikov system is always non-negative while it can be negative in the two-soliton solution of the KP equation.

The collision of two dark-dark solitons is illustrated in Fig.3. It can be seen that after collision, the two solitons pass through each other without any change of shape, darkness or velocity in both components. Therefor, there is no energy exchange between the two solitons or between the $\Phi^{(1)}$ and $\Phi^{(2)}$ components after collision. This complete transmission of energy of dark-dark solitons in both components after collision is a common phenomenon of the coupled Mel'nikov system which occurs for all possible combinations of $\sigma_1$ and $\sigma_2$ values (all-positive, all-nagetive and mixed types).


The reason of the complete energy transmission in dark-dark soliton collisions is that the intensity profile of each dark-dark soliton is completely determined by the background parameters $\rho_i,\alpha_i$ and the soliton parameters $a_i,b_i$. For the colliding solitons, these parameters are the same, hence do not change through collisions. Therefor, the intensity profile of each dark-dark soliton does not change before and after collision.

\section{Dark-Dark Soliton Bound States}
In the study of dark solitons, multi-dark-soliton bound states is a fascinating subject. To get two dark-dark soliton bound states of the coupled Mel'nikov system, the two solitons should have the same velocity in both short and long wave components, so that the two constituent dark solitons can stay together for all times. This means that the parameters need to satisfy
$\frac{\omega_1}{k_{x,1}}=\frac{\omega_2}{k_{x,2}}$ and $\frac{\omega_1}{k_{y,1}}=\frac{\omega_2}{k_{y,2}}$.

\subsection{The stationary dark-dark soliton bound states}
The stationary dark-dark soliton bound states suggest that the common velocity is zero. This requires the coefficients of $t$ in the solution (\ref{i38})-(\ref{i40}) to be zero, i.e.,
\begin{align}
& \frac{\sigma_1 \rho^2_1}{a^2_i+(b_i-\alpha_1)^2}+\frac{\sigma_2 \rho^2_2}{a^2_i+(b_i-\alpha_2)^2}-8(a^2_i-3b^2_i)=0 \ \ \  \textmd{for} \ \ \ i=1,2.
\end{align}
These constraints can be satisfied for all possible combinations of $\sigma_1$ and $\sigma_2$ values. However, for the coupled YO system \cite{chen1} and the coupled NLS system \cite{ohta}, the corresponding constraints are possible only when nonlinearity coefficients take opposite signs. Hence, for the coupled Mel'nikov system, the stationary dark-dark soliton bound states are expected to exist for all possible combinations of nonlinearity coefficients including all-positive, all-negative and mixed types.

Two different types of bound states are illustrated in Figs.4 and 5 respectively. A nontrivial case of $\frac{k_{y,1}}{k_{x,1}} \neq \frac{k_{y,2}}{k_{x,2}}$ is displayed in Fig.4, which corresponds to an oblique bound state. Meanwhile, a trivial case of $\frac{k_{y,1}}{k_{x,1}}=\frac{k_{y,2}}{k_{x,2}}$ is displayed in Fig.5, which corresponds to a
quasi-one-dimensional one.


\subsection{The moving dark-dark soliton bound states}
If the common velocity being nonzero, i.e., $\omega_1\neq 0$ and $\omega_2\neq 0$, we can get the moving dark-dark soliton bound states.
Thus the parameters need to satisfy the following conditions:
\begin{align}
& b_1=b_2,\\
& \frac{\sigma_1 \rho^2_1}{a^2_1+(b_1-\alpha_1)^2}+\frac{\sigma_2 \rho^2_2}{a^2_1+(b_1-\alpha_2)^2}-8a^2_1=\frac{\sigma_1 \rho^2_1}{a^2_2+(b_2-\alpha_1)^2}+\frac{\sigma_2 \rho^2_2}{a^2_2+(b_2-\alpha_2)^2}-8a^2_2.\label{i46}
\end{align}
These constraints are not possible when $\sigma_1$ and $\sigma_2$ are both positive. The reason is that when $\sigma_1=\sigma_2=1$, the function on the left-hand (also right-hand) side of Eq.(\ref{i46}) is a decreasing function of $a^2_j$, thus it can not be satisfied for two different positive values $a_1$ and $a_2$ for the same values $b_1=b_2$. However, when $\sigma_1$ and $\sigma_2$ take opposite signs or they are both negative, the function on the left-hand (also right-hand) side of Eq.(\ref{i46}) may become non-monotone in $a^2_j$, hence it becomes possible for Eq.(\ref{i46}) to admit two different positive values $a_1$ and $a_2$ for the same values $b_1=b_2$.
For the coupled NLS equations and the coupled YO system, the corresponding constraints can be satisfied only when the coefficients of nonlinear terms take opposite signs. To demonstrate these moving dark-dark soliton bound states, the following parameters are chosen
\begin{align}\label{i47}
& \sigma_1=\sigma_2=-1,\ \rho_1=1,\ \rho_2=2,\ \alpha_1=1,\ \alpha_2=\frac{1}{2},\ p_1=1+\textmd{i},\ p_2=0.6012+\textmd{i},\ \xi_{10R}=\xi_{20R}=0
\end{align}
and the corresponding profiles are illustrated in Figs.6-8 at different times.


An important feature of the bound states is that, for both stationary and moving bound states, as $x$ and $y$ move from $- \infty$ to $+ \infty$, the short wave components acquire non-zero phase shifts while the long wave component admits no phase shift. Actually, let $2\phi^{(1)}_j$ and $2\phi^{(2)}_j$ are the phases of constants $K^{(1)}_j$ and $K^{(2)}_j$ respectively, then the phase shifts of the components are $\Phi^{(1)}_{\textmd{phase shift}}=2\phi^{(1)}_1+2\phi^{(1)}_2,\Phi^{(2)}_{\textmd{phase shift}}=2\phi^{(2)}_1+2\phi^{(2)}_2$ and $u_{\textmd{phase shift}}=0$. It can be seen that the total phase shifts of each short wave component are equal to the sum of the individual phase shifts of the two constituent dark solitons, and are generally non-zero, while the phase shifts of the long wave component are always zero. For instance, the total phase shift of the $\Phi^{(1)}$ component is 2$\pi$, and the total phase shift of the $\Phi^{(2)}$ component is -3.9684, as can be calculated from the above formula.

\section{General $N$ Dark Soliton Solutions of the Multi-component Mel'nikov system}
In this section, the previous analysis is extended to the Multi-component Mel'nikov system (\ref{i06})-(\ref{i07}) to get its general $N$ dark soliton solutions. As is known that the multi-bright soliton solutions can be obtained through the reduction of multi-component KP hierarchy, and the multi-dark soliton solutions can be derived from the reduction of single KP hierarchy but with multiple copies of shifted singular points. Consequently, the general $N$ dark soliton solutions of the multi-component Mel'nikov system can be obtained in the same manner as the two-component Mel'nikov system. Here, only the results are presented without a detail calculation.

The multi-component Mel'nikov system (\ref{i06})-(\ref{i07}) consisting of $M$ short wave components and one long wave component can be converted to the following bilinear form
\begin{align}
& (D^2_x+2 {\rm i} \alpha_k D_x-{\rm i} D_y)g_k \cdot f=0,\ \ \ \ \ k=1,2,\cdots,M,\label{i69}\\
& (D^4_x+D_xD_t-3D^2_y)f\cdot f=\sum^M_{k=1}\sigma_k \rho^2_k(f^2_k-hh^\ast),\label{i70}
\end{align}
by virtue of the dependent variable transformations
\begin{align}
& \Phi^{(k)}=\rho_k {\rm e}^{{\rm i}[\alpha_k x+\alpha^2_k y+\beta_k(t)]} \frac{g_k}{f},\ \ \ \ \ k=1,2,\cdots,M,\\
& u=2 (\log f)_{xx},
\end{align}
where $f\equiv f(x,y,t)$ is a real function, $g_k\equiv g_k(x,y,t)$ are complex functions, $\rho_k$ are positive constants, $\alpha_k$ are real constants and $\beta_k(t)$ are arbitrary real functions for $k=1,2,\cdots,M$.

Also introduce a new independent variable $s$, Eq.(\ref{i70}) can be decoupled into
\begin{align}
& (D^4_x+D_xD_s-3D^2_y)f\cdot f=0,\\
& (D_xD_t-D_xD_s)f\cdot f=\sum^M_{k=1}\sigma_k \rho^2_k(f^2_k-hh^\ast).
\end{align}
Consider the Gram determinant solutions of the KP hierarchy, in the similar way, we can get the following $N$-dark soliton solutions of Eqs.(\ref{i06})-(\ref{i07})
\begin{align*}
&f=\Bigg|  \delta_{ij} + \frac{1}{p_i+p^*_j} \textmd{e}^{\xi_i+\xi^*_j}  \Bigg|_{N\times N}, \\
&g_k=\Bigg| \delta_{ij} + \Big(-\frac{p_i-\textmd{i}\alpha_k}{p^*_j+\textmd{i}\alpha_k}\Big)  \frac{1}{p_i+p^*_j} \textmd{e}^{\xi_i+\xi^*_j} \Bigg|_{N\times N},
\end{align*}
with
\begin{align*}
& \xi_j=p_jx -\textmd{i}p^2_jy
 + \frac{1}{2}\Big(\sum^M_{k=1}\frac{\sigma_k\rho^2_k}{p_j-\textmd{i}\alpha_k}-8p^3_j \Big)t  + \xi_{j0},
\end{align*}
where $p_j$ and $\xi_{j0}$ are complex constants, $\delta_{ij}$ is the Kronecker symbol.

\section{Conclusions}
In this paper, the $N$-dark-dark soliton solutions of the coupled Mel'nikov system containing two short wave components and one long wave component are constructed in Gram determinant form by virtue of the KP hierarchy reduction method. The derivation of $N$-dark-dark soliton solutions relies on the connections of the Mel'nikov system with other integrable equations in the KP hierarchy. In addition, the similar expression of general $N$-dark soliton solutions of multi-component Mel'nikov system comprised of $M$ short wave components and one long wave component are also obtained by simply inserting more copies of the shifts of singular points. As far as we know, general $N$-dark soliton solutions of two-dimensional (2D) multi-component soliton equations are rather rare except that the $N$-dark soliton solutions of the 2D multi-component Yajima-Oikawa (YO) systems are studied in Ref. \cite{chen1}.

The dynamics of single and two dark-dark solitons of the coupled Mel'nikov system with two short wave components are studied in detail, and several interesting structures of the solutions have been illustrated through some figures. For single dark-dark solitons, we have shown that the degrees of "darkness" in the two components are different in general. For two dark-dark solitons, in contrast with bright-bright solitons, it has been shown that the collisions of dark-dark solitons are elastic and there is no energy transfer in two components of each soliton. What's more, the dark-dark soliton bound states including both stationary and moving cases are also discussed. For the stationary case, the constraints can be satisfied for all possible combinations of nonlinearity coefficients including all-positive, all-negative and mixed types. For the moving case, the corresponding constraints can be satisfied when they take opposite signs or they are both negative. Whereas, for the coupled YO system \cite{chen1} and the coupled NLS system \cite{ohta}, the dark-dark soliton bound states are possible only under the condition that the coefficients of nonlinear terms take opposite signs.

Recently, the general mixed $N$-soliton solutions of the one-dimensional multi-component YO system and the vector NLS equations are obtained in  Refs. \cite{chen2} and \cite{feng} respectively. Motivated by the remarkable work in Refs. \cite{chen2} and \cite{feng}, the general $N$-bright and $N$-bright-dark solitons of the multi-component Mel'nikov system may also be obtained by the KP hierarchy reduction method. However, the reductions for bright and bright-dark solitons are different from the ones for dark solitons in the current paper. On the other hand, rogue waves, as a special phenomenon of solitary waves originally occurring in the deep ocean, have drawn more and more attentions in many other fields \cite{he1,he2,wen}. As mentioned in Section 1, the KP hierarchy reduction method can also be applied to derive lump and rogue wave solutions of soliton equations. Although the rogue waves of the Mel'nikov system (\ref{i01})-(\ref{i02}) have been investigated in Ref. \cite{qin}, the general rational solutions (lump and rogue wave solutions) of the multi-component Mel'nikov system (\ref{i06})-(\ref{i07}) have not been reported so far (to our knowledge). These questions are interesting and meaningful, we will study them in our future papers.

\section*{Acknowledgment}
We would like to express our sincere thanks to S.Y. Lou, J.C. Chen and other members of our discussion group for their valuable comments and suggestions.
The project is supported by the Global Change Research Program of China (No.2015CB953904), National Natural Science Foundation of China (No.11675054 and 11435005), and Shanghai Collaborative Innovation Center of Trustworthy Software for Internet of Things (No. ZF1213).

\end{document}